SHORT REPORT

# Old vaccines, new usages, surprisingly effective in solving the century-old problem -

## Inactivated African Swine Fever Virus vaccine induces safe and efficient immune protection through mucosal immunity

Yang Jinlong[1*]    Yang JingXu[2*]

([1]Chongqing Academy of Animal Science, Rongchang, Chongqing 402460，China)

（[2]Yeungnam University,Department of global business,Gyeongsan, Gyeongsangbuk-do, 38515，Republic of Korea）

[*] To whom correspondence should be addressed. Tel: +86-23-46792706. Fax: +86-23-46792706. E-mail: yjlfirst@163.com；yang505866726@gmail.com

## Abstract

**Background:** African swine fever (ASF) is among the most devastating viral diseases of pigs. Despite nearly a century of research, there is still no safe and effective vaccine available. The current situation is that either vaccines are safe but not effective, or they are effective but not safe.

**Findings:** The ASF vaccine prepared using the inactivation method with ß-propiolactone provided 98.6% protection within 100 days after three intranasal immunizations, spaced 7 days apart.

**Conclusions:** An inactivated vaccine made from complete African swine fever virus（ASFV）particles using β-propiolactone is safe and effective for controlling ASF through mucosal immunity.

African swine fever (ASF) is an acute, febrile, highly contagious animal infectious disease caused by African swine fever virus (ASFV), which can be transmitted through multiple routes such as the respiratory tract, digestive tract, and skin. The morbidity rate reaches 100%, and the mortality rate is also 100%. It is one of the most severe diseases threatening the pig farming industry [1,2]. Despite nearly a century of vaccine development efforts, due to certain reasons, no safe and satisfactory vaccine has been successfully developed internationally. This has led to a situation where "Vaccines that are safe may not be effective, or vaccines that are effective may pose safety risks". Therefore, developing a safe and effective ASF vaccine has become key to preventing and controlling the spread of this disease [3,4].

Inactivated vaccines, as the most classical method of vaccine development, theoretically represent the safest type of vaccine. However, after injection, ASF inactivated vaccines can only partially stimulate pigs to produce antibodies. Even with the aid of the modern adjuvants, these antibodies cannot effectively neutralize the ASFV. The reason is that these antibodies are specific antibodies, capable only of identification rather than neutralization. Consequently, they fail to achieve the necessary neutralizing effect on the virus, resulting in poor protective efficacy [5]. Therefore, they can only serve as a basis for identifying certain strains of the virus but cannot be used as a vaccine to protect susceptible animals.

ASFV particles primarily spread through mucosal tissues such as the respiratory and digestive tracts during natural infection. Although intramuscular injection of inactivated vaccines does not produce sufficient neutralizing antibodies to protect susceptible animals, our research indicates that mucosal exposure to inactivated whole virus particles can reduce viral load and even clear ASFV. This is likely due to the production of large amounts of SIgA, which can prevent further damage by viral particles at mucosal sites, thereby achieving a preventive effect.

We developed an ASF vaccine using the β-propiolactone inactivation method, which employs complete ASFV particles as antigen components. The envelope proteins remain intact, preserving their immunogenicity, while the viral DNA is fragmented by β-propiolactone, preventing replication (non-spreading, no reversion of virulence), ensuring the safety of the formulation. After mucosal immunization three times to susceptible pigs, precise cross-infection with actively infected ASF pigs was carried out. The results showed a significant reduction or clearance of ASFV loads, achieving excellent preventive effects against ASF.

Our method for preparing inactivated vaccines is as follows:

① Improved sample processing. Sterilely collect liver and spleen tissues from typical African swine fever-infected pigs, remove fat and connective tissue, weigh them, grind thoroughly, mix with 1×PBS buffer solution at a ratio of 1:10, homogenize, freeze-thaw three times, centrifuge at 5000r/min for 15 minutes, take the supernatant, filter through a 0.45μm sterilizing filter, and set aside at 4℃[6].

② viral load testing. ASFV real-time PCR Kit (Beijing MingRiDa Technology Development Co.,Ltd., Beijing, China) was used according to the manufacturer's protocol. The test results showed that the CT value of the prepared samples was less than 30.

③ Preparation of inactivated vaccines. Add 0.1% β-propiolactone to the viral suspension with a real-time PCR detection CT value less than 30, incubate at 4℃ with shaking to inactivate, add another 0.1% β-propiolactone after 24 hours, inactivate for 96 hours, then hydrolyze β-propiolactone at 37℃ for 2 hours, and store at 4℃ [7].

The inactivated vaccine was streaked on blood agar medium and incubated at 37℃ for 72 hours, with no growth of contaminants. Five healthy non-immunized piglets were administered the vaccine intranasally, receiving 1 mL per piglet once every 7 days for three consecutive administrations. During a 30-day observation period, there were no adverse reactions or abnormal manifestations, indicating that the prepared inactivated vaccine has good safety [8].

One hundred and fifty pigs, each weighing about 20 kilograms and tested negative for both ASFV antigen and antibodies, were randomly divided into two groups: an experimental group of 140 pigs and a control group of 10 pigs. The experimental group received inactivated vaccine through mucosal immunity (intranasal administration of 1 ml and oral administration of 3 ml per pig), once every seven days, for three consecutive times. The control group received PBS buffer solution mucosal immunity at a dose of 4 ml per pig as experimental group, also once every seven days for three consecutive times. After three immunizations, all 150 pigs were introduced to a pig farm experiencing an ASFV outbreak. These pigs were then randomly distributed into fifteen pens, with ten pigs in each pen. The pigs were naturally exposed to ASFV by cohabitation with infected pigs, ensuring precise cross-infection. The experiment lasted for 100 days. Every ten days, five pigs were randomly selected respectively from the experimental group and the control group to have their blood samples collected. After collection, the blood was separated into serum, and then the serum from the experimental group and control group were mixed separately before detection. The viral load of ASFV in the blood serum was detected using real-time PCR, and the mortality rate in each pen was recorded. The results, as shown in Table 1, indicate that the viral load in the experimental group of pigs gradually decreased over time and was completely cleared by day 70. At the end of the experiment, 98.6% of the experimental group survived with two pigs dying on days 7 and 50. The surviving pigs did not show any adverse reactions or abnormalities, while the control group had a 100% mortality rate within 25 days.

This study primarily focuses on：(1) The preparation of an inactivated vaccine for ASF using whole virus particles as the antigen and β-propiolactone as the inactivating agent; (2) The method of immunization involves stimulating mucosal immunity through nasal spray and oral administration. After three times of immunization, experimental pigs were cohabited with diseased pigs, and serum viral loads were randomly sampled from the test pigs. The results showed that the viral load gradually decreased with the prolongation of the experiment until it was completely cleared at 70 days, suggesting that the β-propiolactone-inactivated ASF vaccine administered through mucosal immunity may induce high levels of SIgA. In precise cross-infections of ASF, the immune protection rate can reach 98.6% over a period of 100 days.

The results of the experiment indicate that mucosal immunization with ASFV β-propiolactone inactivated vaccine is feasible, safe, and can provide satisfactory immune protection. This trial demonstrates the potential of the prepared vaccine in vaccine production and can reduce mortality caused by ASF. We speculate that mucosal immunization with ASFV inactivated by Binary Ethylenimine（BEI）or low-concentration formaldehyde may also achieve similar results, as long as it does not destroy the viral capsid proteins and envelope proteins but only fragments the viral DNA. It remains to be further verified whether other inactivated virus vaccines that do not achieve protective effects through muscle injection are applicable.

**Acknowledgments**

I would like to thank all animal caretakers and technicians for their excellent work.

**Table 1.Serum samples test results by real-time PCR for the viral load of ASFV**

| Group | 20d | 30d | 40d | 50d | 60d | 70d | 80d | 90d | 100d |
|---|---|---|---|---|---|---|---|---|---|
| **Experimental group** | 23.60 | 25.16 | 28.07 | 28.24 | 30.01 | - | - | - | - |
| **Control group** | 29.99 | ● | ● | ● | ● | ● | ● | ● | ● |

"-": Negative

"●": No sample can be collected